\documentclass{aa}  
\usepackage[utf8]{inputenc} 
\DeclareUnicodeCharacter{226B}{\gg}
\DeclareUnicodeCharacter{03C7}{\chi}
\DeclareUnicodeCharacter{03C3}{\sigma}
\DeclareUnicodeCharacter{03B3}{\gamma}
\DeclareUnicodeCharacter{2212}{-}
\DeclareUnicodeCharacter{2264}{\le}
\DeclareUnicodeCharacter{223C}{\sim}
\usepackage{graphicx}
\usepackage{natbib}
\usepackage{bm}
\usepackage{amsmath}
\usepackage{amssymb}
\usepackage{txfonts}
\usepackage{xcolor}
\usepackage{soul}
\usepackage{float}
\usepackage{listings}
\usepackage{multirow}
\usepackage{tabularx}
\usepackage{comment}
\usepackage[colorlinks=true,linkcolor=blue,citecolor=blue,urlcolor=blue]{hyperref}
\bibpunct{(}{)}{;}{a}{}{,}
\usepackage[normalem]{ulem}
\usepackage[percent]{overpic}

\begin{document} 

   \title{Interlinking internal and external magnetic fields of relativistically rotating neutron stars}

 \author{D. Ntotsikas
          \and
          K.N. Gourgouliatos
          }

   \institute{Universe Sciences Laboratory, Department of Physics, University of Patras, Patras, Rio, 26504, Greece.\\
              \email{d.ntotsikas@upatras.gr, kngourg@upatras.gr}
             }

  \abstract
   {While both the internal and external magnetic field of neutron stars have been studied thoroughly via sophisticated methods incorporating relativistic effects at the exterior and a magnetohydrostatic equilibrium at the interior, or even more complex regimes, the fundamental issue of linking the internal to the external field in a self-consistent way remains yet unresolved. To achieve a realistic depiction of the magnetic field, both the internal and external configurations need to be addressed into a single, simultaneous calculation.}
   {Our aim is to solve for the structure of the magnetic field of a neutron star within the stellar interior, and the relativistically rotating magnetosphere, adhering to barotropic equilibria in the interior and the relativistic force-free condition at the exterior.}
   {We solve the axisymmetric pulsar equation for the magnetosphere and the associated equilibrium equations for the neutron star's interior by employing an elliptic solver  using the method of simultaneous relaxation for the magnetic field inside and outside the star. Appropriate boundary conditions are implemented, at the interior of the star, the light cylinder and the external boundary of the integration domain.}
   {We have found self-consistent solutions corresponding to a variety of combinations of internal and external fields. In all cases the external field satisfies the force-free axisymmetric pulsar equation. The internal field satisfies a barotropic equilibrium and extends to the center of the star. If a toroidal field is included at the interior of the star, then it has either the form of a twisted torus confined within the flux surfaces that close inside the star, or it extends to the magnetosphere, but is contained to the field lines that close within the light cylinder.}
   {This work presents a global solution for the internal and the external field of an axisymmetric rotating neutron star. It is shown that the twist of the internal field affects the external field, by increasing the number of open field lines and eventually the spin-down rate of the star. This effect is far more drastic if the toroidal field, and consequently the poloidal current flowing within the star, is allowed to populate the closed field lines of the magnetosphere, rather than if it remains confined in the star. We further remark that the internal field structure depends on the presence of a twisted magnetosphere: if the twist current is not allowed to flow in the magnetosphere it only occupied a narrow toroid at the interior of the star, whereas if the twist currents are allowed to flow in the magnetosphere the internal toroidal field may occupy a significant volume of the stellar interior. Strong magnetospheric currents may also impact the emission mechanisms, and lead to fluctuations in magnetar spin-down rates, moding and nulling of pulsars, a correlation between angular shear and twist, and the general morphology of the pulsar magnetic field leading to various observational manifestations. The magnetospheric toroidal fields may possibly dissipate, thus the system may switch from global twist to internal twist and consequently exhibit transient behavior.}

\titlerunning{Interlinking magnetic fields of neutron stars} 
  \maketitle

\section{Introduction}
\label{sec:intro}

The magnetic field of neutron stars provides the main route for their observation, either via the generation of coherent radio emission as is the case in rotation powered pulsars \citep{2020PhRvL.124x5101P}, or through its decay \citep{2007PhRvL..98g1101P} or particle acceleration and crust heating \citep{Beloborodov:2009}, which is relevant to thermally emitting X-ray neutron stars. Moreover, explosive and transient events in the form of flares \citep{1979Natur.282..587M,1999Natur.397...41H,2005Natur.434.1098H}, bursts \citep{2018MNRAS.474..961C}, giant pulses \citep{2010A&A...515A..36K}, changes in timing properties \citep{2002ApJ...576..381W,2015ApJ...800...33A,2016MNRAS.458.2088P,2017ApJ...841..126S,2019AN....340..340H,2019MNRAS.488.5251L}, nulling, moding \citep{2013MNRAS.433..445R,lyne2010switched} and even some forms of timing noise \citep{Tsang:2013} are related to greater or lesser extent to their magnetic field. This is mediated either through instabilities and major reconfigurations of its structure \citep{parfrey2012twisting,2013ApJ...774...92P,2016MNRAS.463.3381G,2018ASSL..457...57G,2019PhRvR...1c2049G}, or possibly through electric discharges \citep{2004ApJ...616..439S,2014ApJ...795L..22C} and secular variations of the magnetic field \citep{2012A&A...547A...9P}. 

This central role in the observable properties of neutron stars has motivated thorough studies of magnetic field structure both at the interior \citep{2006A&A...450.1077B,2009MNRAS.395.2162L,2013MNRAS.435L..43C,2014PhRvD..90j1501U,2015ApJ...802..121A,2023PhRvD.108h4006S}, and at the exterior; starting from axisymmetry \citep{contopoulos1999axisymmetric,Uzdensky_2003, goodwin2004idealized, Gruzinov:2005, timokhin2006force}, to three-dimensions \citep{Spitkovsky:2006,kalapotharakos2009three}, using techniques ranging from relaxation to force-free electrodynamics and particle in cell \citep{Philippov_2014,Philippov_2015,Kalapotharakos_2018} and approaches based on machine learning \citep{10.1093/mnras/stac3570,10.1093/mnras/stae192}. Despite the high level of complexity of these studies, there have been limited attempts to address simultaneously both the internal and the external field structure, and this has been done by simplifying the configuration, i.e.~by considering a non-relativistic magnetosphere \citep{glampedakis2014inside,2014MNRAS.445.2777F,akgun2016force,Akgun:2017ggw,Akgun:2018} or a light-cylinder sufficiently far from the surface of the star while focusing on the field structure at the interior and the near zone of the star \citep{2014PhRvD..90j1501U,2015MNRAS.447.2821P}. Thus, the question of the impact of the internal field to the external and vice-versa remains unresolved. 

In this work we address the issue of linking the magnetic field of the interior of the star to the magnetosphere. We solve simultaneously for the internal field so that it obeys  a barotropic MHD equilibrium, while the magnetospheric field is in a relativistic force-free equilibrium. The system contains a toroidal field, which is either of the form of a twisted torus confined within the star, or extends at the magnetosphere in the region of the field lines that close within the light cylinder. Following this approach we provide a global solution for the magnetic field structure of an axisymmetric relativistically rotating neutron star.

The plan of the paper is the following. In section \ref{sec:Mathematical Setup}, we present the mathematical setup and we formulate the equations that describe the system. In section \ref{sec:Numerical Setup}, we present the solution strategy we follow. In section \ref{sec:Results}, we present the simulations and the results arising from the calculations. We discuss the properties of the solutions in section \ref{sec:Discussion}. We present some possible astrophysical applications in section \ref{sec:Applications}. We conclude in section \ref{sec:Conclusions}. 

\section{Mathematical setup}
\label{sec:Mathematical Setup}
Here we present the derivation of the equations that will be solved for the equilibrium of the magnetic field at the interior and exterior of the neutron star. We use cylindrical coordinates $(R,\phi, z)$ and assume axial symmetry throughout the system. The length units are chosen so that the light cylinder $R_{LC} = 1 = c/\Omega$ where $c$ is the speed of light and $\Omega$ the angular frequency of the star.

We express the magnetic field through two scalar functions $\Psi(R,z)$ and $I(R,z)$; as follows:

\begin{equation}
    \bm{B} = \nabla\Psi\times\nabla\phi + I\nabla\phi\, ,
    \label{Magnetic_field}
\end{equation}

 where $\nabla\phi = \hat{\phi}/R$. The function $\Psi(R, z)$ represents the poloidal flux, while $I(R, z)$ is proportional to the poloidal electric current passing through a circular disk parallel to the horizontal plane, whose radius is $R$ and is centered on the axis of symmetry at a distance $z$ from the horizontal plane.
 From Gauss's law, the magnetic field has zero divergence which is satisfied by construction, given the definition of equation (\ref{Magnetic_field}).

Next we will use the above expressions in the appropriate framework to describe the equilibrium at the stellar interior and exterior. 

\subsection{Internal field}

In an MHD equilibrium, the force equilibrium is given by the following expression \citep{2009A&A...499..557R}:

\begin{equation}
    \frac{1}{c\rho}\left(\bm{j}\times\bm{B}\right) = \frac{1}{\rho}\nabla P + \nabla\Phi
    \label{force_balance}
\end{equation}

where $\bm{j}$ is the electric current density, $ P$ is the pressure, $\Phi$ is the gravitational potential, and $\rho$ is the stellar density. Under the assumption of barotropicity,  pressure  $P$ is a function of density $\rho, P = P(\rho)$. Taking the curl of equation (\ref{force_balance}) and using Amp\`ere's law, we obtain:

\begin{equation}     \bm{\nabla}\times\left(\bm{B}\times\frac{\bm{\nabla}\times\bm{B}}{\rho}\right) = \bm{0}\, .
     \label{B_MHD}
 \end{equation}

Since the above expression is a curl of a vector, equal to ${\bm 0}$, we can set the quantity inside curl equal to a gradient of a scalar function $S(R,z)$: 

\begin{equation}
  \nabla S = \frac{1}{\rho}\bm{B}\times(\nabla\times\bm{B})     \, .
    \label{scalar_function}
\end{equation}

From the requirement of axisymmetry, it is evident that:

\begin{equation}
\left[\bm{B}\times(\nabla\times\bm{B})\right]_{\phi} = 0\, ,
    \label{toroidal_component}
\end{equation}

thus any poloidal current must be parallel to the poloidal magnetic field, that leads to the following relation:

\begin{equation}
    \nabla\Psi\times \nabla I = 0 \Leftrightarrow I = I(\Psi) \, ,
    \label{I(Psi)}
\end{equation}

The poloidal component of equation ($\ref{scalar_function}$) satisfies the following equation:

\begin{equation}
  \left(\frac{\partial^2\Psi}{\partial R^2} - \frac{1}{R}\frac{\partial\Psi}{\partial R} + \frac{\partial^2\Psi}{\partial z^2} + I\frac{dI}{d\Psi}\right)\nabla\Psi 
     +  R^2 \rho(r) \nabla S ={\bf 0}\,,
\label{FF_R}
\end{equation}

where we have assumed that the density is a function only of the spherical radius $r = \sqrt{R^2+z^2}$. From equation (\ref{FF_R}), it is evident that the gradient of $S$ is parallel to that of $\Psi$, $\nabla\Psi || \nabla S $, leading to the conclusion that $S = S(\Psi)$. Eventually, the barotropic equilibrium equation is expressed as follows:

\begin{equation}
\Delta^*\Psi + II^{'} + R^2 \rho(r)S^{'} = 0\,,
    \label{Hall_eq}
\end{equation}

where: 
\begin{equation}
\Delta^* =\frac{\partial^2}{\partial R^2} - \frac{1}{R}\frac{\partial}{\partial R} + \frac{\partial^2}{\partial z^2}  \,,
    \label{G-S_operator}
\end{equation}

is the Grad-Shafranov operator and a prime denotes differentiation with respect to $\Psi$.

Regarding the mass density $\rho$ we assume the mass density $\rho$ of the star is approximated through the relation:

\begin{equation}
    \rho(r) = \rho_{0}\frac{r_{ns}^2-r^2}{r_{ns}^2} \, ,
    \label{matter_density}
\end{equation}

where $r_{ns} = 0.1$ is the radius of the star and $\rho_{0}$ is the density at the center of the star. Equation (\ref{matter_density}) corresponds to the solution of the mass density of a neutron star, for an $n=1$ polytrope, where $P=\rho^2$ \citep{2001ApJ...550..426L}.

Despite the functional freedom of $S(\Psi)$, as shown by \cite{2016MNRAS.463.2542G} there are some restrictions on the applicable forms. Here, we assume that $S$ is a linear function of $\Psi$, thus its derivative appearing in the final partial differential equation is a constant  and we set its value $S^{\prime}=1$. This form is compatible with the requirements of the aforementioned work. The choice of a linear form for $S(\Psi)$ restricts the solutions. However, it allows magnetic fields in the form of a dipole for $I=0$ and assuming vacuum external boundary conditions, something that is not possible for non-linear expressions of $S$ in $\Psi$ \citep{10.1093/mnras/stt1195}. Moreover, given the large number of free parameters appearing in the problem, we have chosen to adopt the linear form as a first approach and we will explore the wider parameter space in future work.

\subsection{External field}
\label{Relativistic force-free axisymmetric magnetosphere}
 
Considering the relativistically corotating, plasma-filled magnetosphere, we assume that electromagnetic forces prevail over gravitational, inertial and pressure gradient forces. Therefore, a system in equilibrium must have zero Lorentz force, specifically:

\begin{equation}
    \rho_q \bm{E} + \frac{1}{c}\bm{j}\times\bm{B} = 0\, ,
    \label{ff}
\end{equation}

where $\rho_q$ is the electric charge density. By assuming ideal MHD, the expression of equation (\ref{Magnetic_field}) for the magnetic field and Ohm's law for the electric field corresponding to a corotating magnetosphere, where the rotation axis coincides with the dipole axis, so that:

\begin{eqnarray}
    {\bm E} = -\frac{\Omega R}{c} {\hat \phi} \times \bm{B} 
\end{eqnarray}
we obtain the axisymmetric pulsar equation \citep{scharl_wag}:

\begin{equation}
     (1 - R^2)\left(\frac{\partial^2\Psi}{\partial R^2} - \frac{1}{R}\frac{\partial\Psi}{\partial R} + \frac{\partial^2\Psi}{\partial z^2}\right) -2R\frac{\partial\Psi}{\partial R} = -I(\Psi)I^{\prime}(\Psi)\,,
     \label{FF_magnetosphere}
\end{equation}
where the lengths have been suitably normalized to the light-cylinder radius.\\

Considering the relativistic Grad-Shafranov operator:
\begin{equation}
  \Delta_*= (1 - R^2)\left(\frac{\partial^2}{\partial R^2} - \frac{1}{R}\frac{\partial}{\partial R} + \frac{\partial^2}{\partial z^2}\right) -2R\frac{\partial}{\partial R} 
  \label{relativistic_operator}
\end{equation}
equation (\ref{FF_magnetosphere}) can be written as:

\begin{equation}
     \Delta_*\Psi=  -I(\Psi)\frac{dI(\Psi)}{d\Psi} \, .
     \label{mag_eq}
\end{equation}

In the standard pulsar solution \citep{contopoulos1999axisymmetric}, $I(\Psi) = 0$ in the area of closed field lines, while for the open field it is $I(\Psi) \neq 0$ ensuring a smooth transition through the light cylinder region. Magnetic fields of minimal complexity exhibit a condition of zero poloidal current inside regions of closed magnetic lines, based on the hypothesis that pulsar fields are not inherently twisted. The implementation of a nonzero current $I(\Psi)\equiv I_{tw}\neq 0$ in this area, along with the concept of a twisted magnetosphere, is prevalent in magnetar models and, subsequently, in strongly magnetized systems that will be studied here.  We will refer to this current as the twist current $I_{tw}$ and it will appear in the right-hand-side of equation (\ref{FF_magnetosphere}).

\section{Solution strategy}
\label{sec:Numerical Setup}

\begin{figure}
      \includegraphics[width=.50\textwidth]{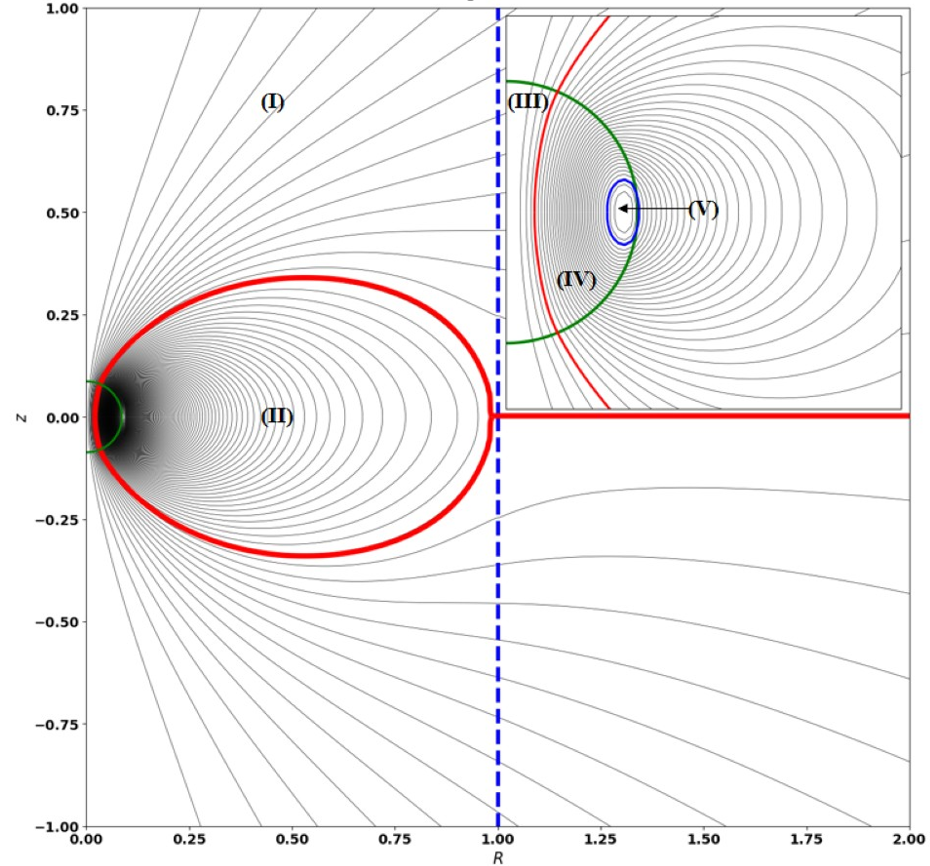}
    \caption{The various regions of the domain, labeled from (I) to (V). The field lines are depicted in black, the last open field line of the magnetosphere in red, the stellar surface in green and the light cylinder in dashed blue. The top right inlet provides a zoom on the star.} \label{fig:1}  
\end{figure}

We study two main families of models.
One, in which the toroidal twist field is allowed to exist only within the star. We will refer to this type of twist as an internal twist. In the second family of models, closed magnetospheric fields lines and field lines at the interior of the star that are connected to the former can be twisted. We will refer to this twist as the global twist, since it extends both at the stellar interior and the magnetosphere. We remark that the open field lines have, in either model, a toroidal field, which is related to the smooth crossing of the light cylinder.

The solution domain is partitioned in five regions, as shown in Figure \ref{fig:1}. Region (I) is the part of the magnetosphere containing open magnetic field lines that cross the light cylinder. Region (II) is the area of the closed magnetic field lines in the magnetosphere containing 
 closed magnetic field lines. Regions (III), (IV), and (V) are all the stellar interior, where region (III) contains field lines that cross the stellar surface and connect to the open magnetospheric field lines, region (IV) contains magnetic field lines that connect to the closed field lines of the magnetosphere, and region (V) contains field lines that close within the star; thus, they are not connected to the magnetosphere and form closed toroids. 

The magnetospheric field in region (I) requires a poloidal return current  $I_{rt}$, essential for the smooth crossing of the light-cylinder. The magnetospheric field in region (II) satisfies the relativistic force-free equation; however, unlike region (I), there is freedom in the form of the current as no field lines cross the light cylinder and thus there is no requirement for a return current. However, it is possible that these field lines are twisted. The magnetic field at the interior, regions (III), (IV) and (V) satisfies equation (\ref{Hall_eq}) corresponding to the barotropic equilibrium. We note that formally, the cross product of the poloidal magnetic field and the electric current in region (III) exert a force in the $\phi$ direction which is required for the spin-down of the star arising from the penetration of the return current into the stellar crust \citep{10.1093/mnras/stz1507}. In our approach, we will neglect the impact of this current in the interior and find its equilibrium through equation (\ref{Hall_eq}) by setting $I=0$. Regions (II), (IV), and (V) may contain a poloidal current as a result of twisting of the magnetic field, $I_{tw}$. In summary, the subsequent relations describe the field in the five aforementioned regions:

\begin{align}
 &  \Delta_*\Psi=  -I_{rt} I_{rt}^{\prime} &\text{(I)} \nonumber\\
  &  \Delta_*\Psi=  -I_{tw} I_{tw}^{\prime} &\text{(II)} \nonumber\\
& \Delta^*\Psi + R^2 \rho(r)S^{'} = 0  &\text{(III)} \label{problem_eq}\\
& \Delta^*\Psi + I_{tw}I_{tw}^{\prime} + R^2 \rho(r)S^{'} = 0  &\text{(IV), (V)}\nonumber
\label{main_eq}
\end{align}

The relativistic Grad-Shafranov operator appears explicitly in regions (I) and (II) of the magnetosphere, whereas all of the interior of the star is characterized by the non-relativistic Grad-Shafranov operator. This issue can be resolved even under the assumption that all areas of the star's magnetic field are solved through the relativistic Grad-Shafranov operator. Given that the stellar surface radius is $r_{ns}=0.1R_{LC}$, the star's rotation exerts a minor influence on the internal structure of the magnetic field, as this would correspond to a maximum correction of $\left(r_{ns}/R_{LC}\right)^2=10^{-2}$.We have confirmed that changes in the magnetic flux function, when the above correction is performed, do not surpass 1\%. Furthermore, the return current is in general much weaker than the twist current, especially if the latter has a remarkable effect on the magnetosphere, thus the assumption of not including it in the calculation of the internal equilibrium does not alter the structure of the internal field.
 
Regarding the boundary conditions that will be implemented, we proceed as follows. We assume that the magnetic flux function, $\Psi$, is zero along the z-axis:

\begin{equation}
    \Psi(R= 0, z) = 0\,.
    \label{boundary_1}
\end{equation}
The value of $\Psi$ at the star's surface is not defined as that of a dipole, but is determined by solving the equation (\ref{problem_eq}) within the star. In this problem, the light cylinder is positioned at a distance of 10 stellar radii ($R_{LC} = 10r_{ns}$). The distance of the light cylinder chosen correspond to an angular velocity of the neutron star of $\Omega = 3000 \, \text{rad/sec}$, indicating that we are formally examining a millisecond pulsar with a spin frequency of approximately $500$~Hz.

The magnetic field is symmetric with respect to the north and south hemisphere; hence, the subsequent boundary conditions are applied at the equator:

\begin{equation}
    \partial_z\Psi(R< R_T, z=0) = 0
    \label{bound.2}
\end{equation}
where $R_T$ denotes the equatorial distance of the outermost closed field line of the magnetosphere. Subsequent to this intersection and along the equatorial plane, the equatorial current sheet of the magnetosphere is characterized by the expression:

\begin{equation}
   \Psi(R\geq R_T, z=0) = \Psi(R_T, 0)
    \label{bound.3}
\end{equation}
Although most works assume that $R_T$ lies exactly on the light cylinder, this point has been argued to be located between $0.8R_{LC}$ and $0.9R_{LC}$ \citep{contopoulos2024pulsar}. When closed magnetic field lines are twisted above a minimum amount, physically acceptable solutions require $R_{T}$ to be placed towards the star \citep{ntotsikas2024twisted}, which is accounted for in our models.

To guarantee the smooth crossing of the field lines through the light cylinder, we impose continuity on the function and its derivative as follows:

\begin{equation}
    \Psi (R = 1^{+},z) = \Psi (R = 1^{-},z)
     \label{bound.4}
\end{equation}
and the regularization condition:
\begin{equation}
    \partial_R\Psi(R=1^{-},z) =  \partial_R\Psi(R=1^{+},z) = \frac{1}{2}I(\Psi)I^{'}(\Psi) \,. 
     \label{bound.5}
\end{equation}

The last equation also relates the value of the $R$-derivative of $\Psi$ to $I$. For the outer region and the open field lines of the magnetosphere, we implement the split monopole boundary conditions to obtain radial behavior at the boundary of the integration grid $z_{max}, R_{max}$. Formally, boundary conditions of this type are applied at infinite distance from the surface of the star, in which case the magnetic lines eventually acquire a radial shape; however, we observed that the application of split-monopole conditions to the boundaries of our grid, compared to a run with twice the size of the numerical domain, does not bring changes in the results, while it significantly increases the convergence speed of the computation. Relation (\ref{bound.5}) enables us to determine the form of $I(\Psi)$ for magnetic flux function values less than $\Psi_0$, where $\Psi_0$ represents the magnetic flux function at the first closed magnetic field line.

We note that a large fraction of the magnetospheric return current enters the star through a current sheet flowing on the separatrix. Such singularities cannot be handled by the finite difference method used here; therefore, we approximate the current sheet using a Gaussian function with a width of $1\times 10^{-3}\Psi_0$ centered at $\Psi = 0.99\Psi_0$, ensuring that the entire return current flows through the open field lines. Unlike the solutions dealing exclusively with the magnetosphere, in our approach, the field emerging from the star is not a pure dipole, thus the maximum flux function may have a different value depending on the twist imposed. To avoid such discrepancies we normalize appropriately the magnetic flux function so that the first closed field line for the $\alpha=0$ model corresponding to $\Psi =1.23$ for direct comparison with the standard solutions \citep{timokhin2006force} and adhere to this normalization for the rest of the solutions. Here, $\alpha$ is a parameter  that determines the ratio of the electric
current distribution to the difference of the flux function $\Psi$ minus its value at the first closed magnetic field line.
 
With regard to the twist current, we will focus on two cases. The globally twisted model where the closed magnetospheric field lines carry poloidal current (region II) connected to the star (region IV):
\begin{equation}
I(\Psi) =
\begin{cases}
    I_{rt}, & \text{(I)}  \\
    \alpha (\Psi(R,z)-\Psi_0), & \text{(II),(IV),(V) } \\
    0, & \text{(III)}\,.
\end{cases}
\label{model1}
\end{equation}
Alternatively, in the internally twisted model, any poloidal current corresponding to an internal poloidal field in the star closes within the star,  thus there is no poloidal current in the closed field lines (region II), nor in the field lines within the star that are connected to the closed magnetospheric field lines (region IV):

\begin{equation}
I(\Psi) =
\begin{cases}
 I_{rt}, & \text{(I)}\\
0,&\text{(II), (III), (IV) }\\
\alpha (\Psi(R,z)-\Psi_{ssf}), & \text{(V)}\,.
\end{cases}
    \label{model2}
\end{equation}
In all cases, $I_{rt}$ is the appropriate return current determined by the smooth light crossing condition. The poloidal twist current in the star is a linear expression given by the flux function $\Psi$ reduced by either its value at last closed field line ($\Psi_0$), or its value at the toroidal loop that touches the stellar surface, for which $\Psi_{ssf}=\Psi(r_{ns},0)$, i.e.~the value of the flux function on the equator of the star. 

Matching of the stellar and magnetospheric solution is made by demanding continuity on $\Psi$ and $I$. The solution in the stellar interior satisfies $\Delta^{*} \Psi+I_{tw}I^{\prime}_{tw}+R^2 \rho S^{\prime}=0$, while the corresponding one in the magnetosphere is $\Delta_{*} \Psi + I_{tw}I^{\prime}_{tw}=0$. The two equations differ on the form of the Grad-Shafranov operator, being in the one case the non-relativistic and in the other the relativistic one and on the term involving the density. We have already commented on the impact of the non-relativistic term of the Grad-Shafranov operator, leading to differences that scale with the square of the ratio of the neutron star radius to the light-cylinder, in which case is $0.01$. The difference due to the density term, is minimized as the density formally drops to zero on the surface of the star, see equation (\ref{matter_density}), thus the same equation is satisfied in either side of the stellar boundary. We note however, that due to discretizations and the cylindrical coordinates we have adopted, the density on the surface has some discontinuities, which, nevertheless do not lead to sharp transitions on the field structure.

In summary, two families of models of magnetic field configurations are developed. The first one, (equation \ref{model1}), corresponds to the following state: regions (II), (IV), and (V) contain a current  associated to the internal twist of the magnetic field lines that also leaks out in the closed magnetospheric field lines, while open magnetospheric lines carry the essential return current from the outer magnetosphere to the star's interior. In the second one, (equation \ref{model2}) there is only a poloidal twist current in the field lines that close within the star, region (V) without crossing the surface of the magnetosphere, while the open field lines, region (I) carry the essential return current. The choice of the linear expression for the twist current is restrictive given the wide parameter space, however we have chosen to limit ourselves to a simple functional form and explore as a main parameter the ratio of the poloidal current to the flux function. The chosen expressions $I = \alpha (\Psi(R,z) -\Psi_0)$ or $I=\alpha (\Psi(R,z) -\Psi_{ssf})$ guarantee that the toroidal field vanishes along this boundary, thus there is no magnetic field discontinuity because of the twist nor an additional current sheet that would have broken the north-south symmetry. A current sheet discontinuity is the one that already exists due to the presence of the return current that closes along the separatrix between the closed and the open magnetic field lines. 

\section{Simulations and results}
\label{results}
We solve equation (\ref{problem_eq}) for different values of the parameter $\alpha$, which determines the current ratio over the magnetic field, applying the simultaneous relaxation method and considering the two cases of global and internal twist. Our simulations are conducted using a FORTRAN 90 framework. As the system is axisymmetric, the functions will depend only on $R$ and $z$. We divide the numerical domain into $R$ and $z$ with a typical resolution of $800\times 800$; nevertheless, we have run in higher resolutions, $1600\times 1600$ to verify the numerical convergence of our results and a change of less than $0.5\%$ was found. The derivatives are computed numerically by finite difference; a central difference scheme is used for the first derivatives, and a three-point stencil is utilized for the second derivatives. The part of the magnetosphere code is based on a refined version of the code of \cite{Gourgouliatos:2019} also used in \cite{10.1093/mnrasl/slad153}, while the part of the interior code has been developed from the first principles for this particular application and has been benchmarked against \cite{10.1093/mnras/stt1195}.

\begin{figure*}
     a\includegraphics[width=.5\textwidth]{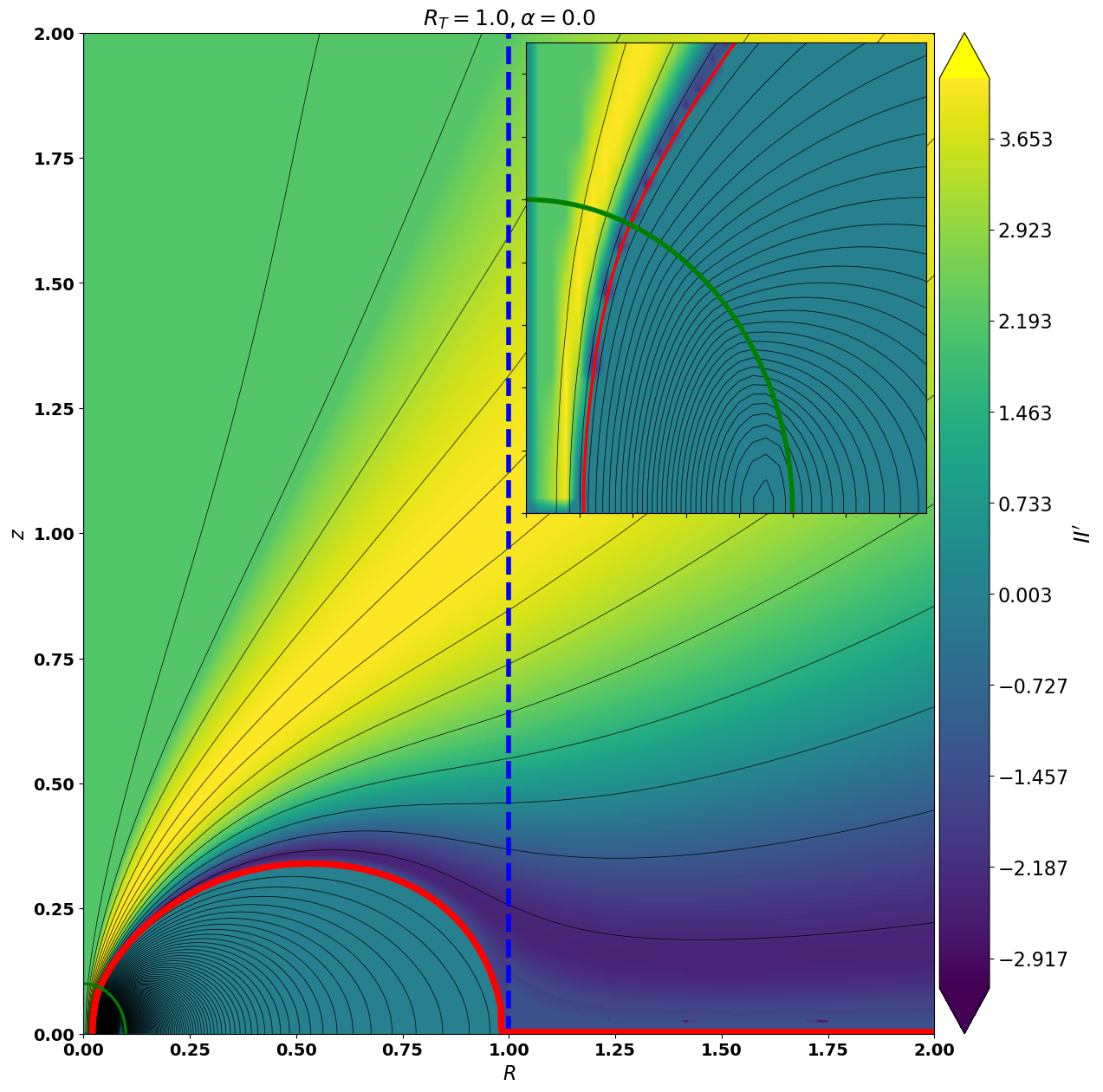}\hfill
     b\includegraphics[width=.5\textwidth]{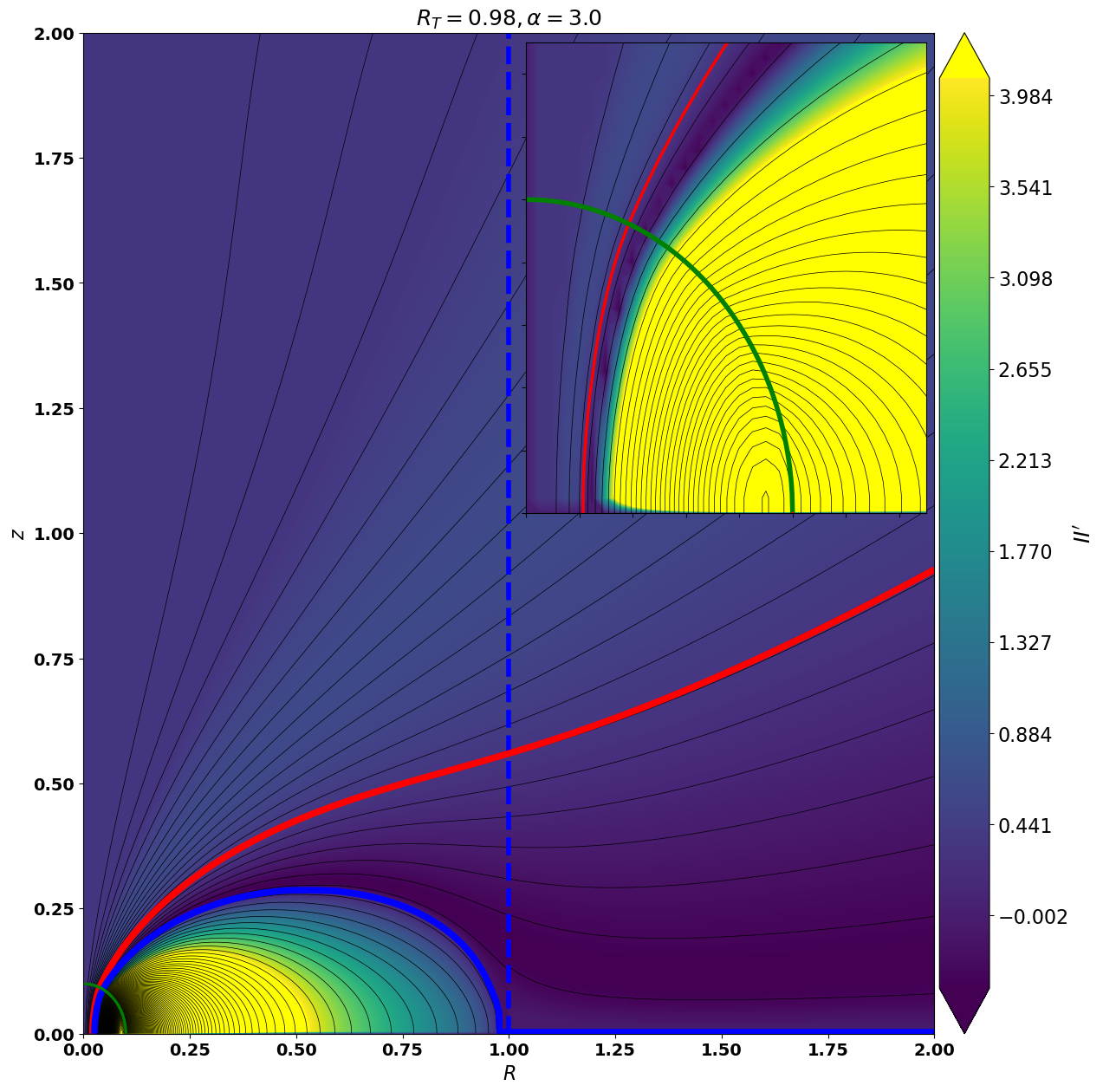}\hfill
    c\includegraphics[width=.5\textwidth]{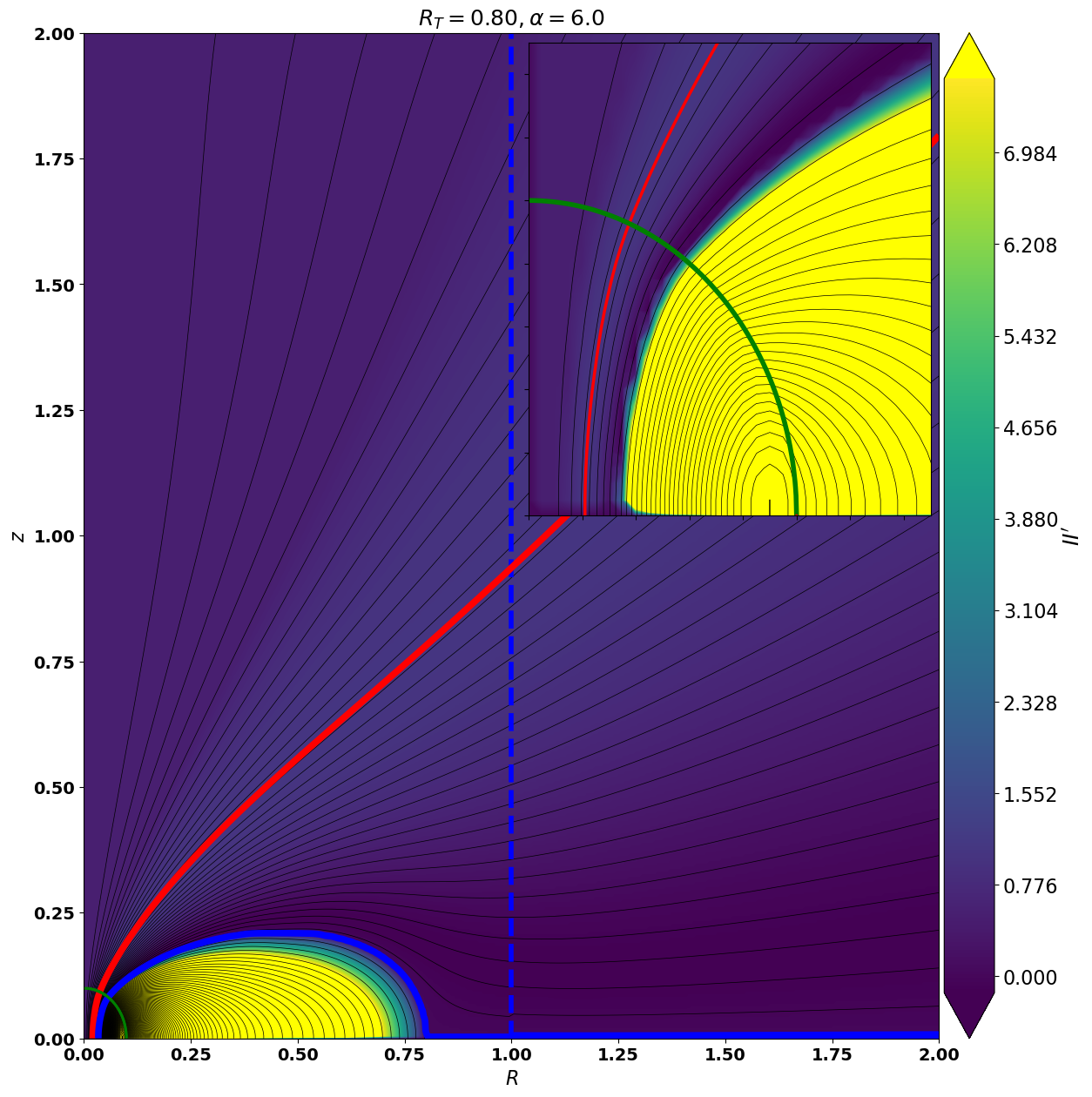}\hfill    
    d\includegraphics[width=.5\textwidth]{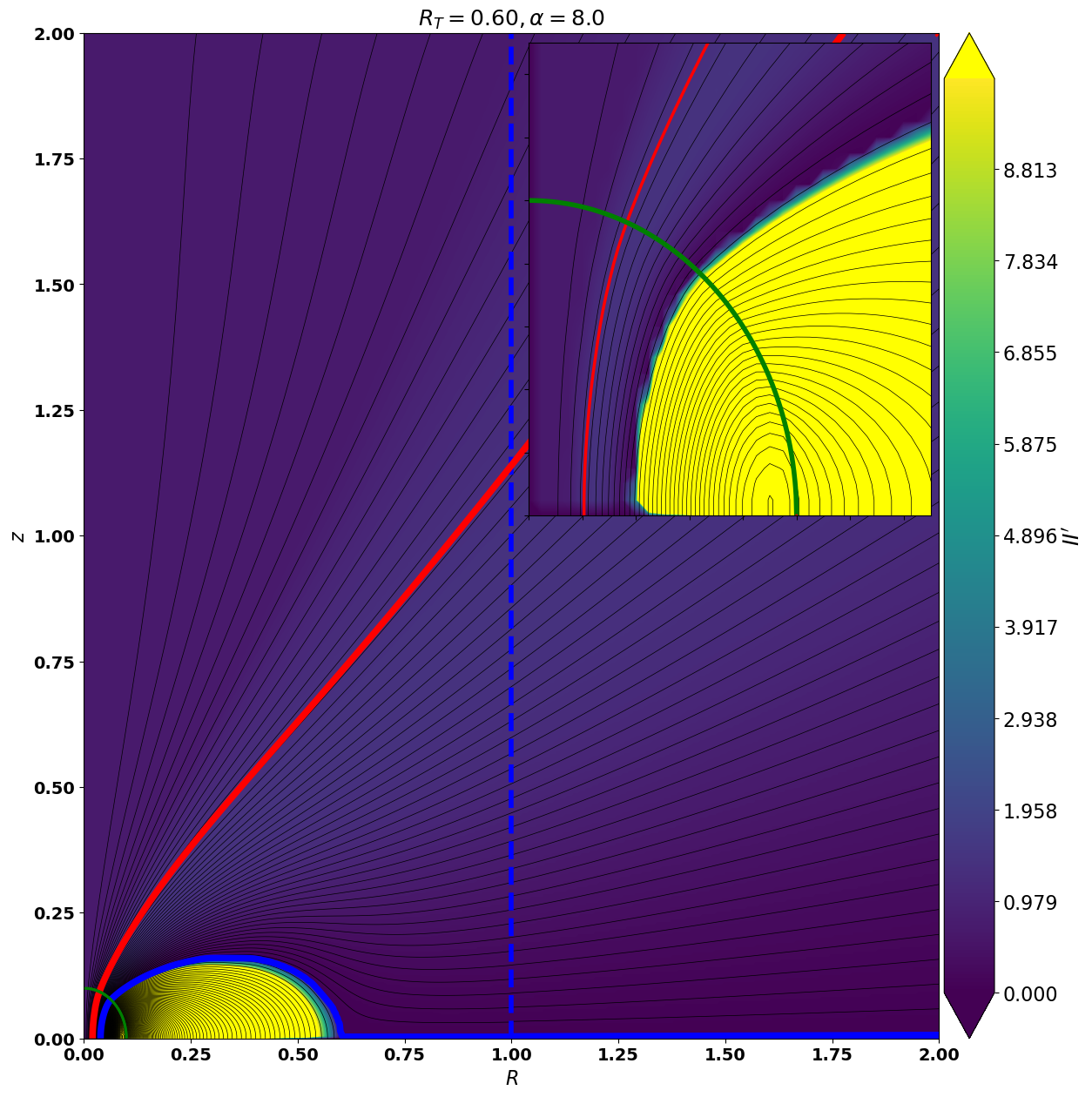}\hfill
    \caption{The magnetic field of a relativistically rotating neutron star and its magnetosphere for the global twist model. The field lines are shown in black, the stellar surface in green, the light cylinder is depicted with the dashed blue line at $R = 1$, along with the current sheet. The red line represents the first closed field line of the untwisted case (panel a) and is shown in subsequent panels for comparison. A poloidal current is applied in the area of the closed magnetic field lines with coefficient  $\alpha = 0.0,~3.0,~6.0,~8.0$. The color bar indicates the value of $II^{\prime}$ across several parts of the star and the magnetosphere. The maximum value of $II^{\prime}$ appearing in the twisted field lines is saturated in color to allow the depiction of its structure in the rest of the system. The top-right inlet is shows a zoomed-in area of the star. } 
    \label{fig:2}
\end{figure*}

\begin{figure*}3     a\includegraphics[width=.5\textwidth]{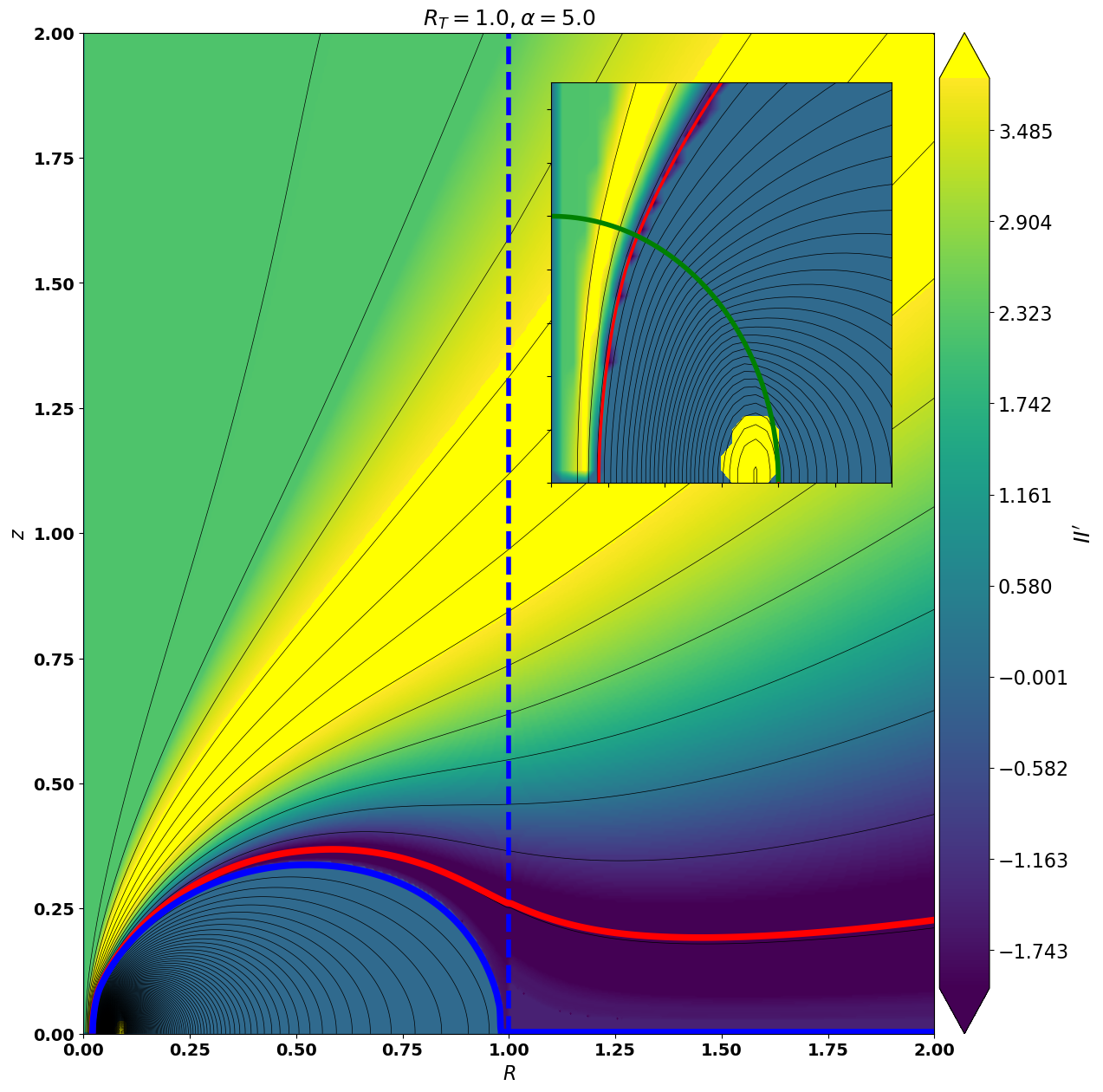}\hfill
     b\includegraphics[width=.5\textwidth]{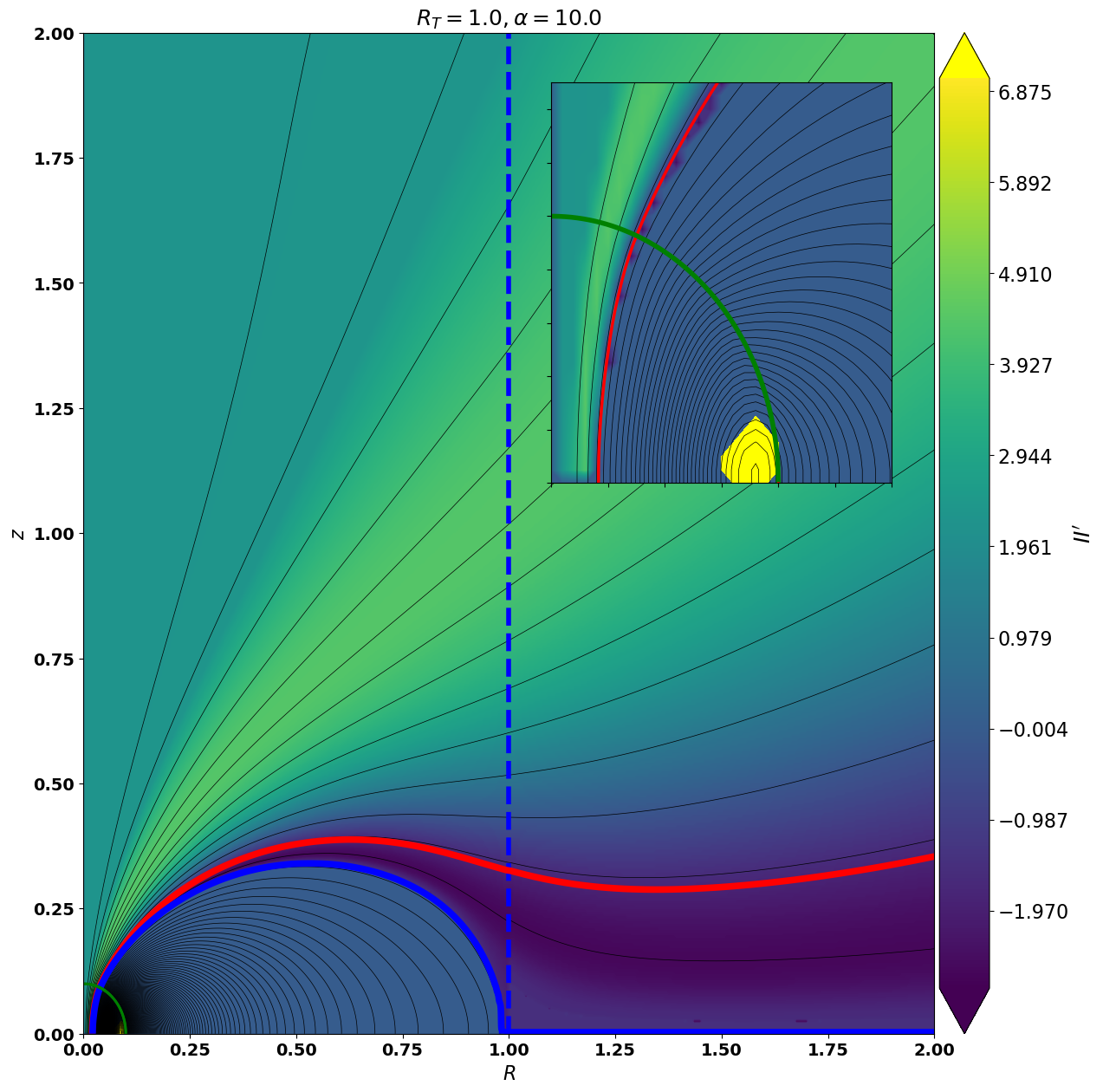}\hfill
     
    \caption{The magnetic field of the internal twist model. Panels a and b correspond to $\alpha = 5.0,~10.0$ respectively. The various field lines, color bars are shown as in Figure \ref{fig:2}. } 
    \label{fig:3}
\end{figure*}

Figure \ref{fig:2} shows magnetic field solutions for the global twist model, for which $I_{tw}$ is applied in the area of the closed magnetic field lines of the magnetosphere, regions (II), (IV) and (V) with coefficients ranging from $\alpha = 0.0$ to $\alpha  = 8.0$. As the parameter $\alpha$ increases, we find that the current of the closed field lines becomes stronger. Additionally, the first closed field line corresponds to a higher value of $\Psi$, and the equatorial current sheet must end closer to the star. We present these effects for various values of $\alpha$ in Table \ref{table:1}, summarizing the main simulation results. Regarding the magnetic field in the interior, we find that the region of the closed field lines shrinks once  more twisted is imposed. We have increased the value of $\alpha$ up to $10$, as beyond this value the twisted region is rather small and under-resolved. Moreover, the innermost point of the equatorial current sheets is twice the radius of the star for the maximum value of $\alpha$ implemented.  

\begin{table} 
\centering
\caption{Simulation results for global twist. }
\label{table:1}
\begin{tabular}{ |c|c|c|c|c|c|c| }
 \hline
 \multicolumn{7}{|c|}{Global Twist} \\
 \hline
 $\alpha$ & $\Psi_0^{\text{norm}}$ & $\frac{L_{\text{twisted}}}{L_{\text{untwisted}}}$ & $\frac{\Delta\phi}{2}$ & $R_T$ & $\Psi_{max}^{norm}$ & $\Psi_{ssf}^{norm}$ \\
 \hline
0.0  & 1.23  & 1.0   & 0.0  & 1.00 & 11.39 & 10.61 \\
\hline
0.5  & 1.25  & 1.03  & 0.13 & 1.00 & 11.40 & 10.62 \\
\hline
1.0  & 1.29  & 1.09  & 0.29 & 1.00 & 11.43 & 10.65 \\
\hline
1.5  & 1.37  & 1.24  & 0.38 & 1.00 & 11.47 & 10.69 \\
\hline
2.0  & 1.49  & 1.47  & 0.44 & 1.00 & 11.53 & 10.76 \\
\hline
2.5  & 1.63  & 1.76  & 0.50 & 1.00 & 11.60 & 10.83 \\
\hline
3.0  & 1.81  & 2.17  & 0.58 & 0.98 & 11.68 & 10.92 \\
\hline
3.5  & 2.02  & 2.7   & 0.61 & 0.95 & 11.77 & 11.02 \\
\hline
4.0  & 2.23  & 3.29  & 0.67 & 0.95 & 11.85 & 11.11 \\
\hline
4.5  & 2.46  & 4.0   & 0.71 & 0.95 & 11.98 & 11.24 \\
\hline
5.0  & 2.73  & 4.93  & 0.79 & 0.90 & 12.10 & 11.36 \\
\hline
6.0  & 3.13  & 6.46  & 0.86 & 0.80 & 12.35 & 11.65 \\
\hline
7.0  & 3.77  & 9.4   & 0.90 & 0.70 & 12.78 & 12.10 \\
\hline
8.0  & 4.10  & 11.10 & 0.94 & 0.60 & 13.10 & 12.44 \\
\hline
9.0  & 4.44  & 13.03 & 0.95 & 0.40 & 13.31 & 12.58 \\
\hline
10.0 & 4.90  & 15.90 & 1.01 & 0.20 & 13.51 & 12.71 \\
\hline
\end{tabular}
\tablefoot{Simulation results for a range of values of the parameter $\alpha$ for the global twist model. The first column is $\alpha$, subsequent columns are $\Psi_0^{norm}$, where $\Psi_0^{norm} =\Psi_0/(7.79\times 10^{-7}) $ , the magnetic flux of the first closed field-line, $L_{twisted}/L_{untwisted}$ the ratio of the spin-down luminosity of a magnetosphere with twist in the closed field-line region $L_{twisted}$ to a standard pulsar solution $L_{untwisted}$, $\Delta \phi$ is a measure of the twist of the closed field-lines, $R_{T}$ the inner radius of the current sheet, $\Psi_{max}^{norm}$ is the maximum value of $\Psi$ attained everywhere in the domain $\Psi_{ssf}^{norm}$ is the maximum value of of $\Psi$ attained on the surface of the star. }
\end{table}

\begin{table} 
\centering
\caption{Simulation results for internal twist.}
\label{table:2}
\begin{tabular}{ |c|c|c|c|c| }
 \hline
 \multicolumn{5}{|c|}{Internal Twist} \\
 \hline
 $\alpha$ & $\Psi_0^{\text{norm}}$ & $\frac{L_{\text{twisted}}}{L_{\text{untwisted}}}$ & $\Psi_{max}^{norm}$ & $\Psi_{ssf}^{norm}$ \\
\hline
0.0  & 1.23 & 1.0  & 11.39 & 10.61 \\
\hline
0.5  & 1.25 & 1.03 & 11.39 & 10.61 \\
\hline
1.0  & 1.29 & 1.09 & 11.40 & 10.62 \\
\hline
1.5  & 1.30 & 1.12 & 11.40 & 10.62 \\
\hline
2.0  & 1.35 & 1.20 & 11.41 & 10.63 \\
\hline
2.5  & 1.41 & 1.31 & 11.42 & 1.064 \\
\hline
3.0  & 1.48 & 1.45 & 11.42 & 10.64 \\
\hline
3.5  & 1.60 & 1.70 & 11.44 & 10.65 \\
\hline
4.0  & 1.68 & 1.86 & 11.45 & 10.66 \\
\hline
4.5  & 1.77 & 2.07 & 11.46 & 10.66 \\
\hline
5.0  & 1.89 & 2.36 & 11.47 & 10.67 \\
\hline
6.0  & 1.94 & 2.49 & 11.50 & 10.71 \\
\hline
7.0  & 1.99 & 2.62 & 11.53 & 10.74 \\
\hline
8.0  & 2.09 & 2.89 & 11.58 & 10.78 \\
\hline
9.0  & 2.29 & 3.47 & 11.61 & 10.80 \\
\hline
10.0 & 2.45 & 3.97 & 11.68 & 10.90 \\
\hline
\end{tabular}
\tablefoot{Simulation results for a range of values of the parameter $\alpha$ for the internal twist model, where the current is confined to the region (V) inside the star. The first column is $\alpha$, subsequent columns are $\Psi_0^{norm}$, the magnetic flux of the first closed field-line, $L_{twisted}/L_{untwisted}$ the ratio of the spin-down luminosity of a magnetosphere with twist in the closed field-line region $L_{twisted}$ to a standard pulsar solution with twist only in the open field-lines $L_{untwisted}$, $\Psi_{max}^{norm}$ and $\Psi_{ssf}^{norm}$ as shown in Table \ref{table:1} We note that $\Delta \phi=0$ and $R_{T}=1$, thus they are not shown here.}
\end{table}

We have also solved for the internally twisted model, Table \ref{table:2} and Figure \ref{fig:3}. As the field is only twisted in the interior, we notice that the twisted region tends to shrink with increasing $\alpha$ as in the globally twisted case. No displacement of the innermost point of the current sheet is required here, as the twist is only at the interior, thus it affects mainly the magnetic flux emanating from the star and its multipolar structure, with higher multipoles appearing for stronger toroidal fields. These effects impact both the flux value of the first closed field line and the structure of the of the magnetic field in the magnetosphere.

We note that for both families of solutions, increasing $\alpha$ leads to a larger amount of flux emerging from the star $\Psi_{ssf}$ and a larger maximum value of flux at the interior of the star $\Psi_{max}$, while all other quantities involved in the equation (density profile and $S$) are kept the same. We note however that the change on either the maximum flux and the total flux emerging from the surface of the star is small and does not scale with the increase of $\Psi_0$. Thus the primary cause of increasing the fraction of open field lines is the change of the magnetosphere. Furthermore, we note that while the form of the flux function on the stellar surface is dominated by a dipolar term, higher multipoles co-exist leading to a different surface magnetic field profile than a pure dipole.   

Comparing the two families of solutions, while the trend is qualitatively the same, we find that the quantitative differences between the two cases are rather pronounced, namely for the same values of $\alpha$ the fraction of the open field lines is substantially higher once twist currents flow through the magnetosphere. This has major impact on various physical properties of the neutron star ranging from the spin-down power to the twist of the magnetosphere and the polar cap opening angle.

\label{sec:Results}

\section{Discussion}
\label{sec:Discussion}
Next we are going to address the consequences of increasing the twist to the main physical properties of neutron stars. We will focus on the spin-down luminosity, the overall twist in the magnetosphere, and the polar cap opening angle.

\subsection{Spin-down Luminosity}

\begin{figure}
     \includegraphics[width=.49\textwidth]{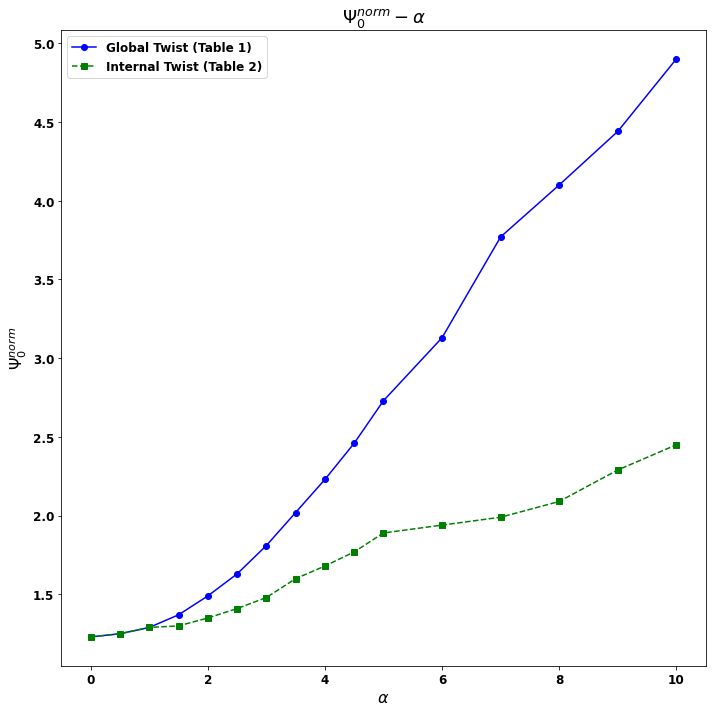}
    \caption{Magnetic flux function of the first closed magnetic field line as a function of $\alpha$.} 
    \label{fig:4}
\end{figure}

\begin{figure}
     \includegraphics[width=.49\textwidth]{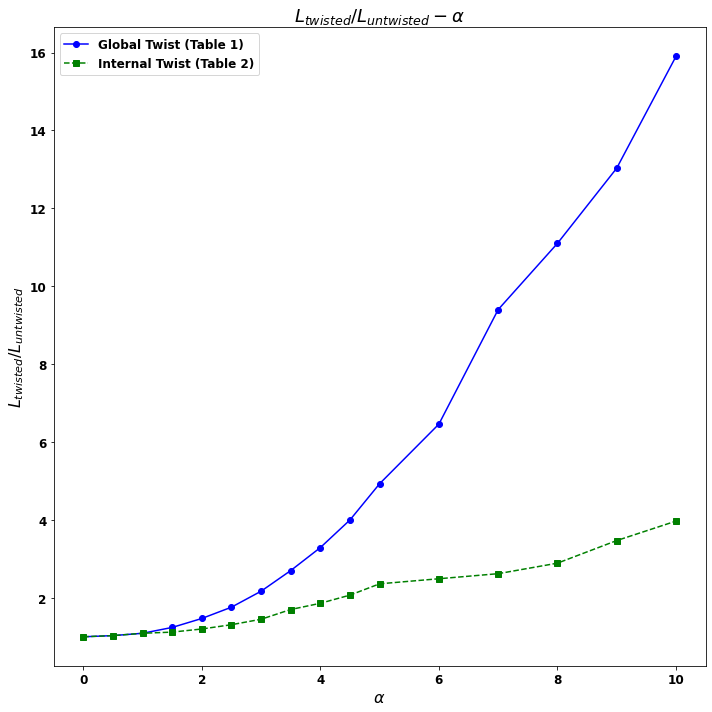}
    \caption{The spin-down luminosity of the various solutions scaled to the spin-down power of the untwisted solution, as  function of $\alpha$.} 
    \label{fig:5}
\end{figure}

The general trend is that an increase in the twist current leads to a larger fraction of open field lines. For small values of $\alpha\leq 1$, both models give the same result, Figure \ref{fig:4}, however, once $\alpha$ exceeds this value, the global twist model has a much higher value of $\Psi_0$. This is also clearly visible in Figures \ref{fig:2} and \ref{fig:3} where the field line first closed of the untwisted model shown in red is much more drastically displaced from the equator in the globally twisted model than in the internally twisted one.  

The overall spin-down luminosity, which corresponds to the loss of electromagnetic energy from the star's magnetic field, is expressed by the following relation: 

\begin{equation}
    L = 2\int_0^{\Psi_0^{norm}}I(\Psi)d\Psi
    \label{luminosity1}.
\end{equation}
Using equation (\ref{luminosity1}), we evaluate the star's spin-down power. In models where the twist current flows in the magnetosphere, the spin-down power can be enhanced by a factor of approximately $16$ for $\alpha=10$, Figure \ref{fig:5}, whereas if the twist current closes within the star, for the same value of $\alpha$ the spin-down luminosity is enhanced by a factor of $4$. The rapid increase of the spin-down luminosity in the globally twisted model is related to the fact that the current sheet inner edge is closer to the stellar surface; therefore there is a drastic increase of the open magnetic field lines.

\subsection{Magnetospheric twist}

Next, we evaluate the twist of a magnetic field line by integration of the following relation:

\begin{equation}
    \frac{dR}{B_R} = \frac{dz}{B_z} = \frac{Rd\phi}{B_{\phi}}
    \label{twist}
\end{equation}
where $B_R$ is the radial component of the magnetic field, $B_z$ the axial and $B_{\phi}$ the azimuthal. Thus, we obtain the following:
\begin{equation}
    \Delta\phi = \int\frac{B_{\phi}}{RB_R}dR \,.
    \label{deltafi}
\end{equation}
As the twist may vary in different field lines, i.e.~the fist closed field lines have zero twist, as defined there by $B_{\phi}=0$, while the field line emerging from the equator of the star has the maximum value of $I$, however its extend is formally zero. Therefore, we perform this integration is evaluated on the field line that corresponds to $\Psi = 1.1\Psi_0^{norm}$. We integrate from the surface of the star to the equator; therefore, the results presented in Table \ref{table:1} represent half the twist values that would be derived by measuring the whole extent of the magnetic field line, emerging from the northern and closing at the southern hemisphere. In agreement with earlier solutions \citep{lynden1994self,gourgouliatos2008fields,pavan2009topology,Akgun:2017ggw}, the twist tends to approach a maximum value $\sim \pi/2$ rather than increasing indefinitely. Therefore, the system can formally approach a split monopole state after a finite amount of twist. We have decided to investigate the properties in terms of the parameter $\alpha$ up to a maximum value of $\alpha=10$ in our calculation.

\begin{figure}
     \includegraphics[width=.49\textwidth]{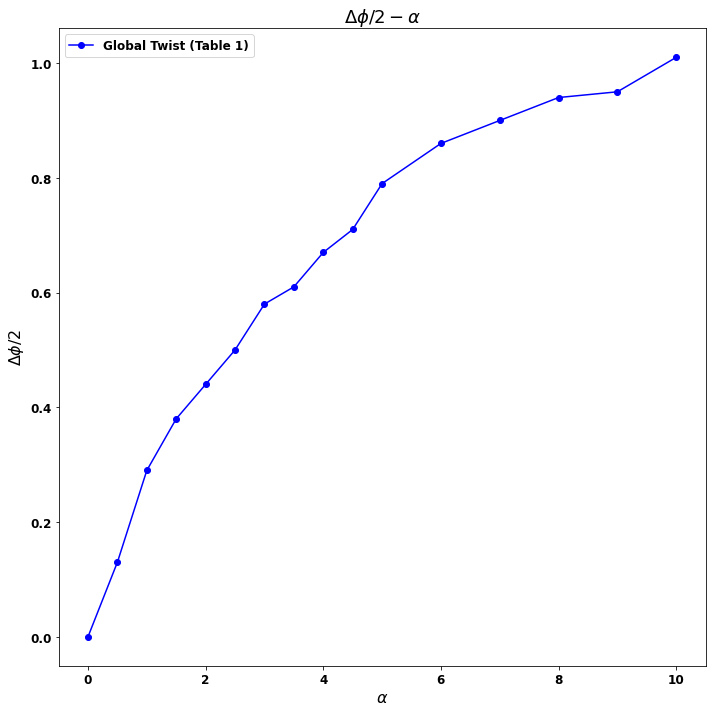}
    \caption{The twist of the magnetospheric part of the field line for which $\Psi=1.1\Psi_0$, as a function of the parameter $\alpha$.} 
    \label{fig:6}
\end{figure}

Figure \ref{fig:6} shows the twist of the closed magnetic field line region. It is evident that with increasing $\alpha$, the twist in region (II) becomes higher; however, this region shrinks and is concentrated closer to the star, and the twist at the exterior will eventually vanish.

\subsection{Polar Cap}

We calculate the semi-opening angle of the polar cap by tracing the foot points of the first closed field line corresponding to $\Psi_0$. The location $(R_{\Psi_0}, z_{\Psi_0})$ of its foot points satisfies the following conditions: $\Psi(R_{\Psi_0}, z_{\Psi_0})=\Psi_0$ and $R_{\Psi_0}^2+z_{\Psi_0}^2=r_{ns}^2$. Then, the semi-opening angle is given by the following expression:

\begin{equation}
     \theta_{pc} = \arctan\frac{R_{\Psi_0}}{z_{\Psi_0}}\,,
     \label{33}
\end{equation}

\begin{figure}
 \includegraphics[width=.49\textwidth]{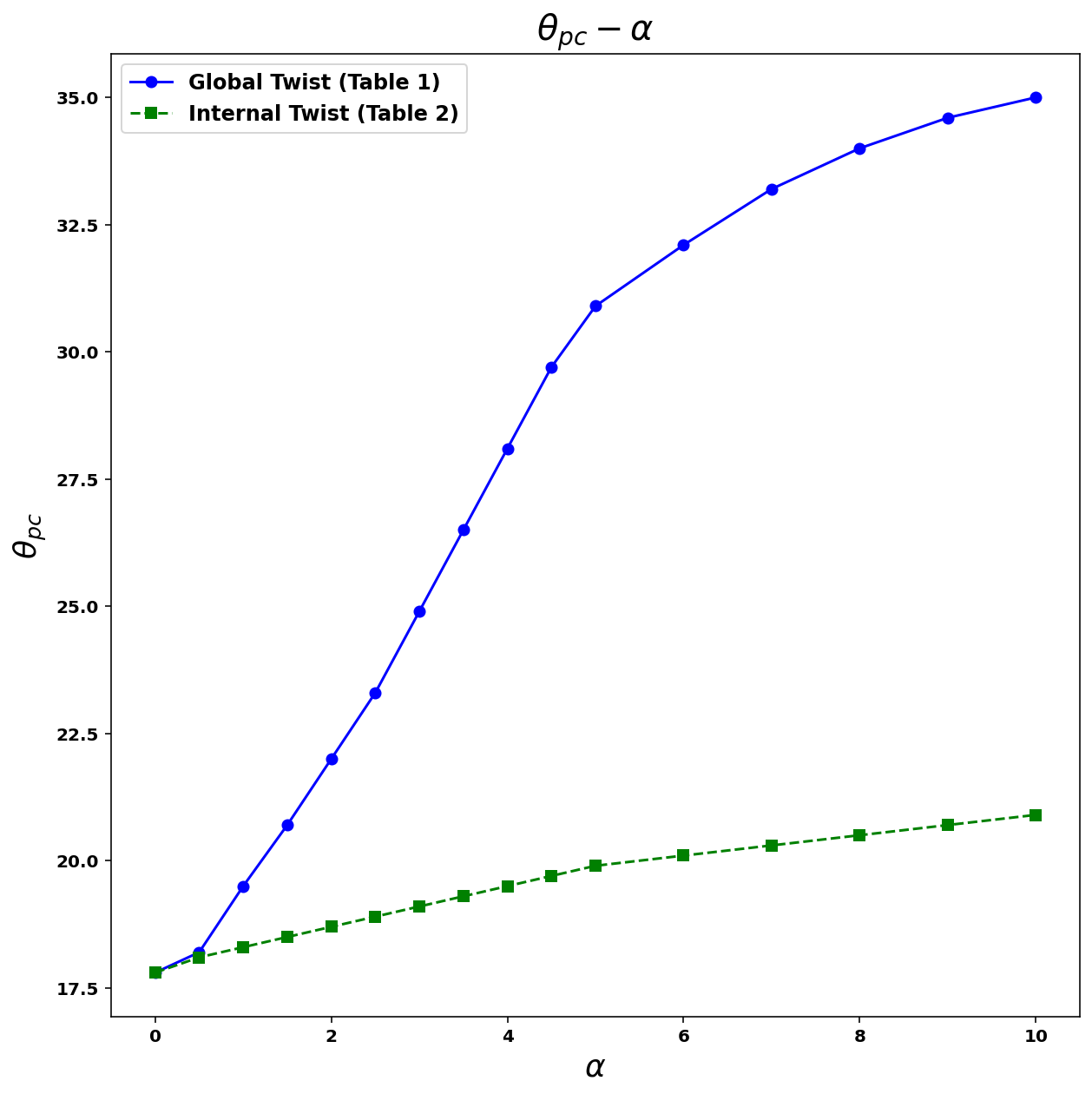}
    \caption{The polar cap's semi-opening angle as a function of $\alpha$. } 
    \label{fig:7}
\end{figure}

 We plot the semi-opening angle of the polar cap in Figure \ref{fig:7}. We find that as $\alpha$ increases, the semi-opening angle of the polar cap expands both for the internal and global twist. However, there is a difference in magnitude. If the twist current is allowed in the magnetosphere, the polar cap increases very rapidly, and it even becomes a split monopole after some finite twist. In contrast, if the twist current lies entirely within the star and does not populate the closed magnetospheric field lines, changes on the polar cap are minor, increasing from $17.5^{\circ}$ to $20^{\circ}$, due to the fact that the magnetic field on the surface is no longer dipolar, but it contains higher multipoles of north-south symmetry.
\section{Applications}
\label{sec:Applications}
The solutions found here illustrate the impact of twist on the global magnetic field structure of a neutron star, which spans from the interior to the exterior. Our models explore the major qualitative difference of the presence of twist in the magnetosphere or exclusively at the interior of the star, and employ an incremental exploration of the parameter space regarding the injection of twist of the field lines. These differences in the magnetic field structure both at the interior and the exterior are likely to impact the observational behavior of neutron stars. 

Indeed, we confirm that the inclusion of a toroidal magnetic field strongly affects the spin-down rate. Even if a comparable amount of poloidal magnetic flux crosses the stellar surface, a twisted field will have a more pronounced spin-down rate. We notice major differences in the spin-down rate depending on whether the twist current is confined within the star or whether it populates the magnetosphere. For smaller values of $\alpha$ up to $\alpha\approx 1$, the spin-down power scales the same for both families of models. For values higher than that, we find that there is a drastic difference. In particular, if we allow the twist current to flow in the magnetosphere, for $\alpha=10$ the spin-down luminosity is about $16$ times higher than in the untwisted case, whereas if the twist current is enclosed in the star it is about $4$ times higher compared to the untwisted model. This effect will differentiate the two families of models. This is related to the fact that the equatorial current sheet needs to retract closer to the star if the magnetosphere is twisted, whereas the current sheet can start at the light cylinder in the absence of external twist. This calculation provides further evidence that the twist of the magnetosphere enhances the spin-down luminosity of the star \citep{2002ApJ...574..332T,2006MNRAS.367.1594L,2011A&A...533A.125V}, but also quite remarkably that internal twist also leads to spin-down luminosity, even if it only indirectly affects the magnetosphere by altering the lower boundary conditions.

We note that our calculations are constrained by the fact that the light cylinder is only ten times larger than the stellar radius. A smaller star or a more distant light cylinder would allow for a finer exploration of the innermost point of the current sheet and would not stop at this particular value. In the highly twisted case, where the currents flow in the magnetosphere, we further notice that the polar cap becomes larger and the magnetic field is reminiscent of a split monopole even within the light-cylinder, thus the field lines are much less curved in the poloidal direction, yet there is still curvature due to the toroidal field. This stretching of the magnetic field disfavors the generation of curvature radiation, as a smaller part of the field is curved, as opposed to the previous case.  

A critical question is what determines whether the internal twist will also populate the magnetosphere, or if it is only going to be contained in the star, in principle the capacity of the magnetosphere to support twist currents. Twist currents can become much stronger than the spin-down current which is related to the ratio of the neutron star radius to the light cylinder. They require a larger population of charges in the magnetosphere which is related to the multiplicity of the charges and may untwist within timescales of years \citep{beloborodov2007corona}, thus providing a transient, but not necessarily explosive behavior. As this question is still unresolved, we speculate that magnetospheres with a higher charged particle density are more likely to host such currents, whereas magnetospheres with smaller particle densities are possibly reminiscent of vacuum or minimally twisted magnetospheres. This will have the following consequences: neutron stars with twisted magnetospheres will have a higher spin-down rate, the field lines will have a larger curvature radius, and, as postulated, they will have a higher charge density. This could provide an interpretation of transient magnetar behavior, as the spin-down rate is higher, the generation of curvature radiation related to radio emission is less likely due to the structure of the field lines, and the presence of charged regions and currents within the magnetosphere may eclipse any radio emission produced near the star \citep{2019MNRAS.488.5251L,2023ApJ...945..153L}. We note that such variations on the spin-down power have been noted in simulations of twisted magnetospheres \citep{parfrey2012twisting,ntotsikas2024twisted} and also in the timing behavior of pulsars correlated with emission across the electromagnetic spectrum and changes in pulse profile \citep{2006MNRAS.370L..76U,2007ApJ...663..497C,2008ApJ...679..681C,2020ApJ...904L..21Y,Lower:2025pry,2024MNRAS.528.3833F}, also accompanied by spectral evolution \citep{2025arXiv250220079Y}. 

These models also introduce current sheets that extend within the light cylinder, especially for highly twisted systems. As current sheets are prone to instabilities, they may undergo tearing modes, which lead to dissipation. This effect may be leading to the release of energy and particle acceleration. Although such effects are likely to occur in pulsars \citep{2017ApJ...850..142C,2021A&A...656A..91C,2023ApJ...958L...9B}, this model brings them within the light cylinder and very close to the star. The creation of current sheets is intimately related to the twisting of the external field, thus, torque variations due to switches from twisted and untwisted states will be accompanied by a rapid release of energy through explosive events \citep{2020ApJ...889..160A}. Moreover, the twist, internal or global, affects the size of the polar cap, with larger polar caps associated with higher twist \citep{2019MNRAS.489.3769T}, as has been suggested to be the case in sources with flaring episodes \citep{2020ApJ...889L..27Y}.

In our simulations, the chosen ratio of the light-cylinder radius over the neutron star radius corresponds to a millisecond pulsar whose frequency is approximately $500$~Hz. Although this is much faster than any known magnetar, it may be directly comparable to millisecond magnetars \citep{2022ASSL..465..245D}. These sources are newborn magnetars, expected to spin-down rapidly \citep{2020MNRAS.496.2183C}. In addition to these sources, an interesting extension would be the scaling of the model to less rapidly rotating sources. In such systems, the light cylinder lies much farther away than the 10 stellar radii we have assumed here, and it is possible that a smaller fraction of the closed field lines will be twisted. However, the inclusion of twist may impact the spin-down properties of lower rotating sources, if it extends all the way to the light cylinder. However, in models where the twist current is contained within the star, part of the change in the spin-down power is due to the fact that the field has a multipolar structure on the surface rather than a dipole. This will not affect significantly the field structure near the light cylinder, as higher multipoles decrease rapidly. Nevertheless, even if these effects are mild, a toroidal field affects the overall amount of magnetic flux emerging from the star, thus leading to a higher spin-down rate. 

A further remarkable difference between the models in which the twist is contained in the star versus the ones where the magnetosphere is twisted is the volume occupied by the toroidal field within the star. In the former case, the toroidal field is confined in a very small region, within closed loops of magnetic flux; on the contrary, if the twist is allowed to populate the magnetosphere, it reaches much higher latitudes up to approximately $45^{\circ}$, for the models presented here. This expansion of the toroidal field region can have important implications for the magnetic field energy, as the toroidal field can contribute more intensely to the total energy budget, even under equilibrium conditions, whereas in cases the twisted field was contained in the star this ratio was rather low \citep{2009MNRAS.395.2162L}, unless a special form of the poloidal current was postulated \citep{2013MNRAS.435L..43C}. It may also affect the ellipticity of the star, as the toroidal field tends to increase it, and in the fully confined models it can only contribute locally, whereas in this family of models, the toroidal field can be much stronger and extend to a larger region within the star \citep{2008MNRAS.385..531H}.   

The question of a hydromagnetic equilibrium of a neutron star with rotation was also posed in the work of \cite{glampedakis2014inside}. In that work the problem was studied in the non-rotating system, while touching upon the question of rotation, without, however, fully addressing the light-cylinder and the relativistic magnetosphere. Our results provide a direct comparison between the two regimes. The obvious difference is in the very structure of the magnetosphere, where the additional physical constraints of the relativistic force-free magnetosphere have been added. In particular, the current flowing along the open field lines is determined by the requirement that the magnetic field crosses smoothly the light cylinder, therefore, solutions where a predetermined current is allowed to flow are not acceptable. The presence of the light-cylinder adds an extra length-scale which, in combination with the twist-current, leads to contraction of the closed field line region and the presence of a current sheet closer to the star. Furthermore, the open field lines, especially near the light-cylinder and beyond have a structure of a split monopole: such a field is significantly different from the dipole background assumed there. Notably, split monopole structures can appear in spherical geometry only after a relatively small amount of twist \citep{lynden1994self,gourgouliatos2008fields,2019MNRAS.489.3769T} Apart from the expected differences in the magnetosphere, where the actual equilibrium equation is different, there are further, remarkable differences at the interior. In particular, the internal toroidal field is allowed in the equatorial and mid-latitude region, but not near the poles as in the non-rotating solution of \cite{glampedakis2014inside}. Consequently, the region where the twist current flows, at the interior this time, can maximally be, that of the closed field lines, a region determined by the relativistic solution. Overall, the solution with a relativistically rotating magnetosphere is more constrained than the non-rotation inside-out magnetosphere and provides direct estimation of astrophysically relevant quantities for neutron stars including their relativistic magnetosphere.

\section{Conclusions}
\label{sec:Conclusions}
This research investigates equilibrium configurations for twisted magnetic fields of neutron stars considering both the star and the magnetospheres. Previous studies, in general, either examined the external magnetosphere without accounting for the stellar interior or presumed a vacuum exterior. Our work highlights a comprehensive method, in which the magnetic field is consistently solved for, including the stellar and magnetospheric currents. 

The study indicates that the inclusion of twist impacts  the structure of the neutron stars magnetic field . Its effect is moderate if the twist is only allowed in the interior of the star and rather drastic if the twist is also in the closed magnetospheric field lines. We have quantified its impact on the spin-down rate, magnetospheric twist, and polar cap opening angle, offering insights into phenomena such as transitions between pulsar and magnetar states or intermittent sources. This work has provided the framework for the extraction of observable parameters and the eventual quantification of these effects.

Future research should progress to solutions beyond that of a barotropic equilibrium and simple linear forms for $S(\Psi)$ and the twist current $I_{tw}$. Moreover, it is essential to adopt more realistic models in which both the crust and the core are accounted for, as the relevant equilibrium equations are rather different. Furthermore, progress from axisymmetric models to three-dimensional simulations may provide a more precise understanding of magnetospheric processes.

An in-depth examination of the communication between the star's inside and exterior, particularly over its surface, is crucial. These endeavors will be essential for addressing fundamental inquiries regarding the evolution of magnetic fields in neutron stars and comprehending the relationship between magnetospheric distortions and observable phenomena, such as timing variations in magnetars and even normal sources.

\section{Data availability statement}
\label{sec:Data Availability Statement}
The data sets generated during the current study are available from the corresponding author upon reasonable request.
\begin{acknowledgements}
We are grateful to an anonymous referee whose constructive comments improved this paper. We are grateful to I. Contopoulos and S. Lander for discussions during the preparation of this manuscript. 
This work was supported by computational time granted by the National Infrastructures for Research and Technology S.A. (GRNET S.A.) in the National HPC facility - ARIS - under project ID
pr017008/simnstar2.
\end{acknowledgements}
\bibliographystyle{aa}
\bibliography{bibliography}
\end{document}